\documentclass[prl,floatfix,twocolumn]{revtex4}
\usepackage{graphicx}%
\usepackage{dcolumn}
\usepackage{amsmath}
\begin{document}
\title[Short Title]{Synchrony-optimized Networks of Non-identical Kuramoto Oscillators}
\author{Markus Brede}
\affiliation{%
CSIRO Centre for Complex Systems, GPO Box 284, Canberra, ACT 2601, Australia
}%

\date{\today}
\begin{abstract}
\noindent In this letter we discuss a method for generating synchrony-optimized coupling architectures of Kuramoto oscillators with a heterogeneous distribution of native frequencies. The method allows us to relate the properties of the coupling network to its synchronizability. These relations were previously only established from a linear stability analysis of the identical oscillator case. We further demonstrate that the heterogeneity in the oscillator population produces heterogeneity in the optimal coupling network as well. Two rules for enhancing the synchronizability of a given network by a suitable placement of oscillators are given: (i) native frequencies of adjacent oscillators must be anti-correlated and (ii) frequency magnitudes should positively correlate with the degree of the node they are placed at.
\end{abstract}
\maketitle
An interesting feature of populations of coupled oscillators is synchronization \cite{Winfree}. Questions about properties of coupling schemes that favor synchronization have received much attention in the recent literature \cite{Nishi,Hong,Hwang,Motter,Motter1,Lee,Korniss,Chavez,Donetti,MB,Gleiser}. Most of this work has so far concentrated on systems of identical oscillators and an analysis of the stability of the synchronized state via the eigenvalue ratio \cite{Pecora1} and correlations between it and properties of the coupling network. Typical findings are that network topologies with stable synchronized states tend to be very homogeneous in degree- and load distributions, very small and characterized by having very few short loops and a high degree of disassortative mixing between node degrees \cite{Donetti,MB}. Indeed, very recent studies characterized the most synchronizable networks as `superhomogeneous' \cite{Donetti}. However, such networks are rarely found in nature. Part of the homogeneity of the networks might be attributable to the fact that all nodes are subject to identical dynamics. Variance in the oscillator populations is far more likely in nature and seems to be the more relevant case in many practical applications. Inhomogeneity in the optimal coupling arrangement is very likely to arise from the inhomogeneity in the oscillator population. Thus, understanding the interplay between the native frequency of an oscillator and its place in the coupling network is a very interesting problem.

One step in this direction has already been made in Ref. \cite{Gleiser}, where networks of  non-identical Kuramoto oscillators are rewired by favoring connections between oscillators with similar average frequencies. This study, however, does not investigate the connection between the native frequencies of oscillators and the topology of the optimized networks. The latter is the main focus of  this letter, where some first results about the  optimal coupling of a heterogeneous oscillator population are discussed.

The eigenvalue ratio, which is used in most of the previous work, is a measure of the linear stability of the synchronized state of coupled identical oscillators. Though some network properties that were derived from the eigenvalue anlysis seem to also give rise to better synchronizability for non-identical oscillators \cite{Chavez}, the method itsself can make no statement about the non-identical oscillator case. To address the latter issue, we drop the generality of the eigenratio analysis and concentrate on a specific oscillator dynamics, given by the Kuramoto model \cite{Kuramoto}
\begin{align}
 \label{Kuram}
 \dot{\phi_i}=\omega_i+\sigma \sum_j a_{ij} \sin(\phi_j-\phi_i),
\end{align}
where the $\phi_i$, $i=1,...,N$ describe the phases of $N$ oscillators, which have native frequencies $\omega_i$ and are coupled with strength $\sigma$ via a network described by its adjacency matrix $A=(a_{ij})$. For simplicity we only consider symmetric unweighted coupling. Our choice of the dynamical model is motivated by the simplicity of the Kuramoto model and the fact that it can also be understood as an approximation to a more complex dynamical situation \cite{Kuramoto}.

To measure the degree of synchronization of the oscillators, one frequently introduces the order parameter
$r(t)=|1/N \sum_j \exp(i\phi_j(t))|$ \cite{Kuramoto}. For perfect synchronization, one has $r=1$, whereas $r\sim O(N^{-1/2})$ for randomly drawn omega's and small coupling. A second order phase transition at some critical coupling strength $\sigma_\text{c}$ between a desynchronized and a synchronized phase is found for many coupling schemes \cite{Kuramoto}. For global coupling (see e.g. \cite{Hoppenstead}) and  recently for homogeneous and heterogeneous network topologies \cite{Gardenes,Gardenes1} the route  to synchronization has been well studied.

The synchronizability of a network is characterized by the dependence of $r$ on the coupling strength $\sigma$. A partial ordering might be defined, by calling a network $A_1$ more synchronizable than a network $A_2$, if $r_{A_1}(\sigma)>r_{A_2}(\sigma)$ for all $\sigma$. If $r_{A_1}(\sigma_1)>r_{A_2}(\sigma_1)$ for one $\sigma_1$, but $r_{A_1}(\sigma_2)<r_{A_2}(\sigma_2)$ for another $\sigma_2$ no statement can be made.

In what follows, we are interested in the arrangement of the couplings $a_{ij}$ that leads to the best synchronizability for a fixed number of links and a given set of native frequencies $\{\omega_i\}_{i=1}^N$, where the frequencies $\omega_i$ are drawn from a uniform distribution over $[-1,1]$\cite{remark0}. To address this issue we propose a simple optimization scheme. After chosing a value $\sigma^*$, we start with an Erd\'os-R\'enyi random graph (ER r.g.) coupling \cite{ER}. To determine a network's synchronizability, we then numerically integrate Eq.'s (\ref{Kuram}) for the initial condition $\phi_i(t=0)=0$  and calculate the average value of $r$ over $\Delta T$ integration steps $\overline{r}(\sigma^*)=1/\Delta T \sum_{t=T_\text{rel}}^{T_\text{rel}+\Delta T}r(t,\sigma^*)$, having first allowed for $T_\text{rel}$ timesteps for relaxation. $T_\text{rel}$ is chosen such that average trajectories $\langle r(t>T_\text{rel})\rangle_\omega$ are stationary for the initial network.

Then, at each iteration of the optimization, we suggest a rewiring in the coupling network's configuration. By suggesting a move of a randomly chosen link to some randomly chosen link vacancy, the connectivity of the network is preserved. A new (connected) configuration is accepted, if its average fitness $\overline{r}(\sigma^*)$ is larger than that of the old configuration and rejected otherwise. The procedure is repeated till no move was accepted during the last $I=2\sum_{i,j}a_{i j}$ iterations.

\begin{figure} [tbp]
\begin{center}
\includegraphics [width=7.3cm]{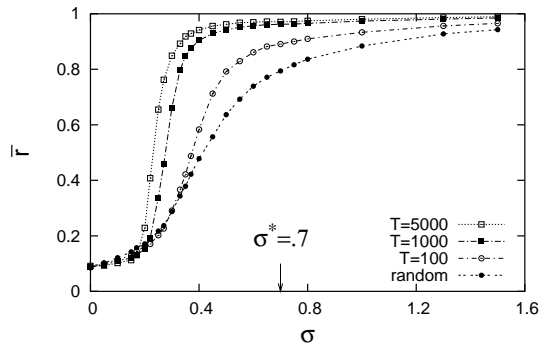}
\caption{Comparison of the dependence $r(\sigma)$ for random and several partially optimized networks (the label $T$ gives the number of optimization steps performed). Averages are performed in the following way: first a set of frequencies is uniformly drawn from $[-1,1]$. Then, for this specific set of omega-values, averages over ensembles of initial conditions randomly drawn from $[-\pi,\pi]$ are calculated (numerically over 10 realizations of initial conditions). In a second step, averages over 100 different sets of native frequencies are performed. The networks contain $N=100$ nodes and have connectivity $\langle k\rangle=3.5$. For the optimization we chose $\sigma^*=.7$.}
\label{F0}
\end{center}
\end{figure}

Three observations justify this somewhat arbitrary procedure. First, we find that for a sufficiently large choice of $\sigma^*$ a network that has superior synchronizability for homogeneous initial conditions $\phi_i(t=0)=0,i=1,...,N$ also has superior average synchronizability if initial conditions are drawn from a uniform distribution over $[-\pi,\pi]$. The reason for this is that the system (\ref{Kuram}) loses memory of the initial conditions for large enough coupling. Second, we note that networks which  are better synchronizable for one (large enough) choice of $\sigma^*$, also prove to be better synchronizable at least for coupling strengths larger than $\sigma^*$ and generally also for a range of coupling strengths below $\sigma^*$. Even more, if for two configurations $A_1$ and $A_2$ for which $r_{A_1}(\sigma^*)>r_{A_2}(\sigma^*)$, configuration $A_1$ will generally also have an earlier onset of synchronization (cf. Fig. \ref{F0}). Third, it is found that optimal networks obtained for different values of $\sigma^*>\sigma_\text{c}$ are very similar. Taking averages over an ensemble of initial networks, we indeed find almost identical $r(\sigma)$-dependencies for ensembles of networks optimized for such different values of $\sigma^*$.  All three observations allow to conclude that the networks obtained from the above optimization procedure have a superior synchronizability in the sense defined by the above ordering.
Contrary to the above, a choice of $\sigma^*$ much smaller than the critical coupling strength for ER r.g.'s only leads to more synchronizable networks for the specific choice of initial conditions, but does not entail enhanced synchronizability if averaged over an ensemble of initial conditions. The reason for this appears to be that memory of the initial state is kept during the evolution of the oscillator system and hence the optimization generates networks that are specifically adapted to a particular initial distribution of phases. Best convergence was generally obtained for choices of $\sigma^*$ such that $r(\sigma^*)\approx.9$ in the initial network.

We seeded the optimization procedure with different initial networks, ranging from scale-free networks constructed after the Barab\'asi-Albert model \cite{Barab}, ER r.g.'s and (almost) regular random graphs. In all cases we find very similar optimal configurations which have almost identical $r(\sigma)$-dependences. Thus, though probably no global optimum configuration is reached, the similarity of the optima clearly shows how networks change towards more synchronizable configurations.

Although in its aim very similar, our optimization procedure differs in some essential points from the one used in \cite{Gleiser}, where rewirings that favour connections between elements with similar average frequencies are performed. The latter may favor partial synchronization, explaining the large level of clustering and the relatively small improvements in $\overline{r}$ for large coupling strength reported in \cite{Gleiser}.

Refs. \cite{Gardenes,Gardenes1} discovered that the way to synchronization in different networks can be very different. Whereas in heterogeneous topologies the fully synchronized state tends to grow from a unique core synchronization cluster around hubs, the transition to full synchronization in homogeneous networks manifests itself by the coalescence of many small synchronization clusters of similar size. In heterogeneous networks the onset of synchronization arises for smaller coupling, but this is traded-off against a lesser stability of the fully synchronized state and a smaller degree of synchronization $\overline{r}$ for larger couplings. By the choice of a large $\sigma^*$ in the balance between the onset of synchronization and stability of the synchronized state more emphasis is put on the latter.

\begin{figure} [tbp]
\begin{center}
\includegraphics [width=7.3cm]{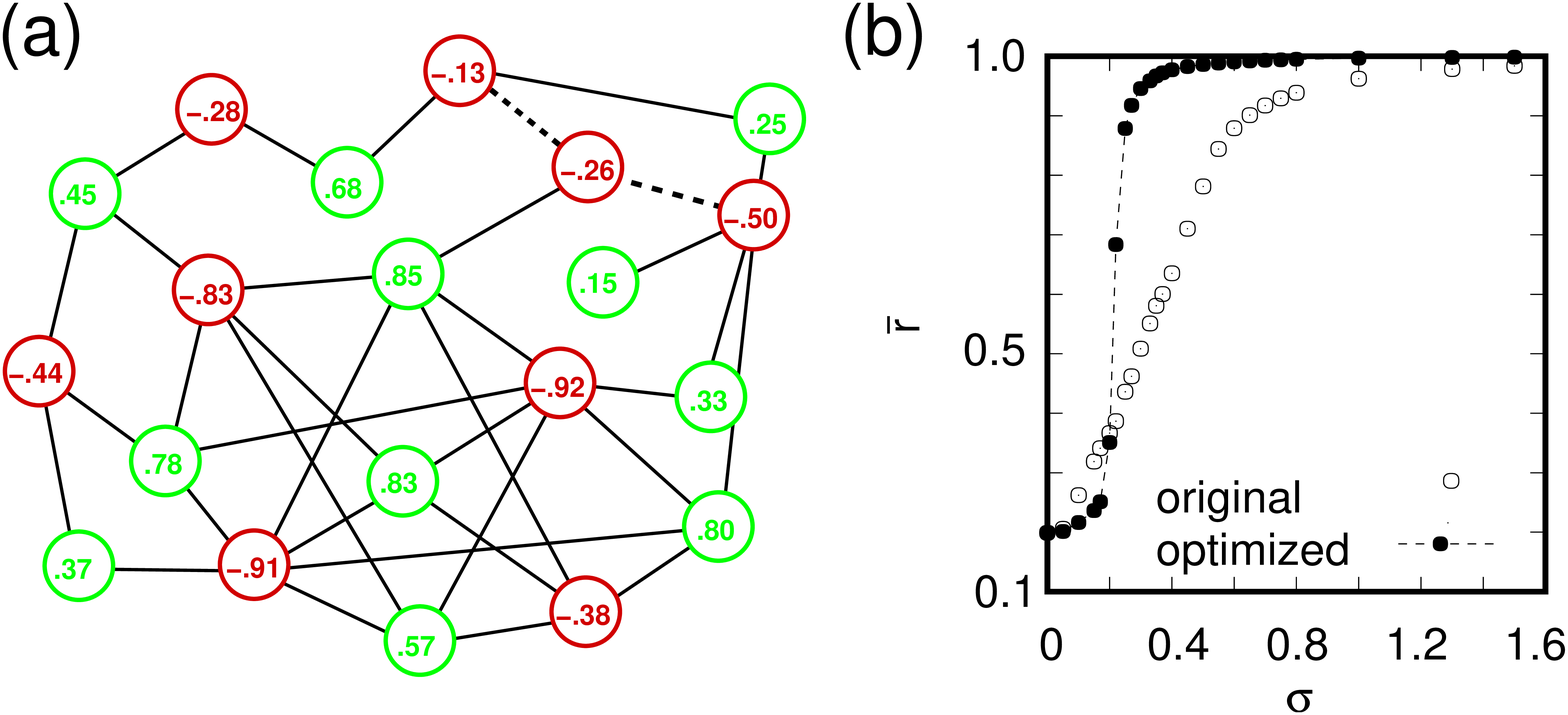}
\caption{Structure of an optimized network (a) and comparison of the sigma-dependence of the degree of synchronization $r$ between the original (open circles) and the optimized network (filled circles) (b). In the network illustration, nodes associated with negative native frequencies are drawn in red, others in green. Connections between negative-positive omega pairs are drawn as solid lines, others as fat dashed lines. Note, that there are only two cases where nodes with frequencies of the same sign are neighbours.}
\label{F1}
\end{center}
\end{figure}

Below we proceed by investigating correlations between network configurations and native frequency arrangements generated by the optimization approach. 
To gain intuition, Fig. \ref{F1}a displays a small optimal network and the corresponding assignment of native frequencies. Closer inspection of the optimal network strongly suggests that its structure is characterized by a pairing of the native frequencies, such that nodes with $\omega>0$ are always surrounded by other nodes with $\omega<0$ and vice versa. We introduce two quantities to describe these correlations between the $\omega$-values of adjacent nodes. First, $p_-$ measures the proportion of links connecting nodes with omegas of different signs. Second, a correlation coefficient
\begin{align}
 c_\omega =\frac{\sum_{i,j} a_{i j}(\omega_i-\langle \omega\rangle) (\omega_j-\langle \omega\rangle)}{\sum_{i,j} a_{i j}(\omega_i-\langle \omega\rangle)^2},
\end{align}
where $\langle \omega\rangle=1/N\sum_i\omega_i$
additionally quantifies how often nodes with large negative omega are adjacent to nodes with strongly positive omega. For arbitrary assignments of native frequencies one has $\langle p_-\rangle=1/2$ and $\langle c_\omega\rangle=0$.

We proceed by investigating the dependence of $p_-$ and $c_{\omega}$  as well as of several network characteristics on synchronizability. For this, we determine averages of the clustering coeficient, pathlength, diameter, assortativeness, load
\begin{align}
b_i=\sum_{j,k} d^{(i)}_{j,k},
\end{align}
where $d^{(i)}_{j,k}=1$ if the shortest path from node $j$ to node $k$ passes through node $i$ and $d^{(i)}_{j,k}=0$ otherwise,
load and degree variance and maximum load during the network optimization. In the following, we limit the discussion to changes in the network, where the starting point of the optimization was a random graph. Starting with other initial conditions, e.g., scale-free networks, changes in most network properties stem from a reorganization of the hub structures towards more `democratic' arrangements. However, in the last stages of the evolution, trends are very similar to those we discuss for ER r.g.'s below.

\begin{figure} [tbp]
\begin{center}
\includegraphics [width=7.8cm]{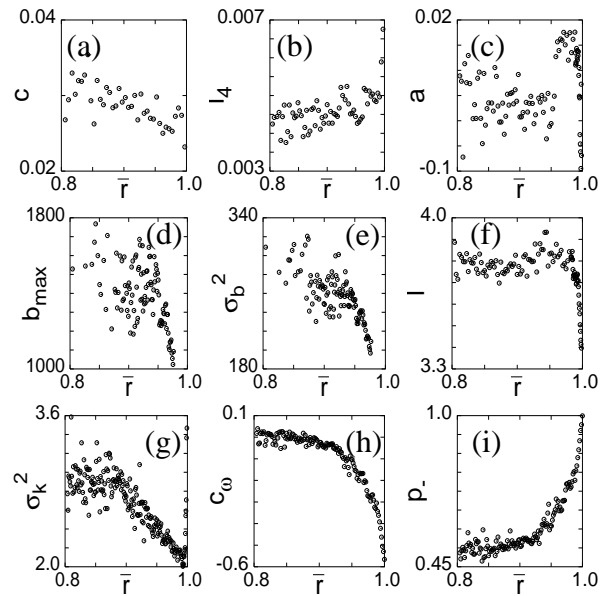}

\caption{Dependence of the clustering coefficient (a), density of 4-loops (b), assortativeness (c), maximum load (d), load variance (e), average pathlength (f), degree variation (g), fraction of plus-minus pairs (h) and of the correlation of adjacent native frequencies (i) on synchronizablity measured by $r(\sigma^*)$ during the optimization. The data are for a network of 100 nodes with connectivity $\langle k\rangle=3.5$. The initial network corresponded to an ER r.g., for other initial conditions like scale-free networks or regular random graphs similar trends are found.}
\label{F2}
\end{center}
\end{figure}

Our most important finding is the observation that the optimal networks are characterized by a strong anti-correlation of adjacent positive and negative frequencies. Both, the number of plus-minus pairs $p_-$ as well as the anti-correlation are found to increase strongly during the evolution of the networks towards better synchronizability (cf. Fig. \ref{F2}h,i).
Further, on the one hand, the changes in the assortativeness, cliquishness, load, maximum load and load variance as well as  in the average pathlengths and diameter (not shown) corroborate results obtained from the eigenratio analysis for identical oscillators. Heterogeneity in the oscillator population  does not appear to affect network properties that are related to the transmission of information. On the other hand, the network configurations with optimal synchronizability are not found to be the most homogeneous ones, and while exhibiting fewer triangles, are not generally marked by lower densities of short loops than random networks. Since triangles always involve at least one adjacent pair of oscillators with native frequencies of the same sign, the decay in the triangle density, and indeed the density of odd loops generally, during optimization follows from the increase in the number of frequency-anticorrelated node pairs. Hence the triangle density can not be used as an indicator for general short loop densities. For this reason, we also measured the density of loops of length four, which is indeed found to increase during the optimization (cf. Fig. \ref{F2}b).

Next, we discuss the development in the networks' degree variance during their evolutions. For this, it appears worthwhile to observe that the largest degree of anticorrelation between adjacent native frequencies can be realized in very inhomogeneous networks, where frequencies of very large (or very small) magnitude attract more links than frequencies of average magnitude. This lets one understand that degree homogeneity and enhanced anti-correlation of adjacent node frequencies are conflicting demands.

Figure \ref{F2}g displays a minimum in the dependence of the degree variance on synchronizability. Initially, the degree variance is reduced, but around $r=0.98$ this trend is reversed. We hypothesize that this point is associated with an arrangement of the network where no further increase in the strength of anti-correlations between adjacent oscillators can be found in very regular coupling schemes. Instead, the degree of anticorrelation can only be further enhanced by placing native frequencies of large absolute value on nodes of higher degrees, which gives them access to many neighbours of low frequency magnitude. As oscillators synchronize with the mean frequency $\langle \omega\rangle\approx 0$ (for even distributions of the $\omega's$), this placement can be understood as resulting in a strong drag on large frequencies towards the mean frequency. The same effect in the correlations could be achieved by placing frequencies of low magnitude on the higher degree nodes and surround them by frequencies of large magnitude. However, the latter arrangement clearly favours the build-up of small clusters of oscillators synchronizing with frequencies of large magnitude.

In fact, the optimal network displayed in Fig.\ref{F0}a appears very close to the first arrangement of the native frequencies. For example, both frequencies of the largest magnitude $\omega=-.92$ and $\omega=-.91$ are assigned to the two nodes with the most connections, while the next two belong to nodes with the second most connections, etc. We also measured the dependence of the average magnitude of the native frequency on the node degree $\langle |\omega| \rangle (k)$ and find a steady increase from $\langle |\omega| \rangle (1)\approx.15$ for nodes of degree one to $\langle |\omega| \rangle (8)\approx.75$ for the highest degree node of the optimal network. Corresponding with this, the average magnitude of the difference between the frequency of a node and the average frequency of its nearest neighbours $\Delta \omega_\text{n.n.} =|\omega-1/k_i \sum_i a_{i j} \omega_j|$ grows from $\langle \Delta \omega_\text{n.n.}\rangle(1)\approx.7$ to  $\langle \Delta \omega_\text{n.n.}\rangle(8)\approx1.08$. Clearly, the optimal network configuration for heterogeneous oscillators is not `superhomogeneous'. Instead `hub' nodes are found which are distinguished by native frequencies of large magnitude, whereas low degree nodes are typically associated with frequencies of relatively small magnitude.

\begin{figure} [tbp]
\begin{center}
\includegraphics [width=4.2cm]{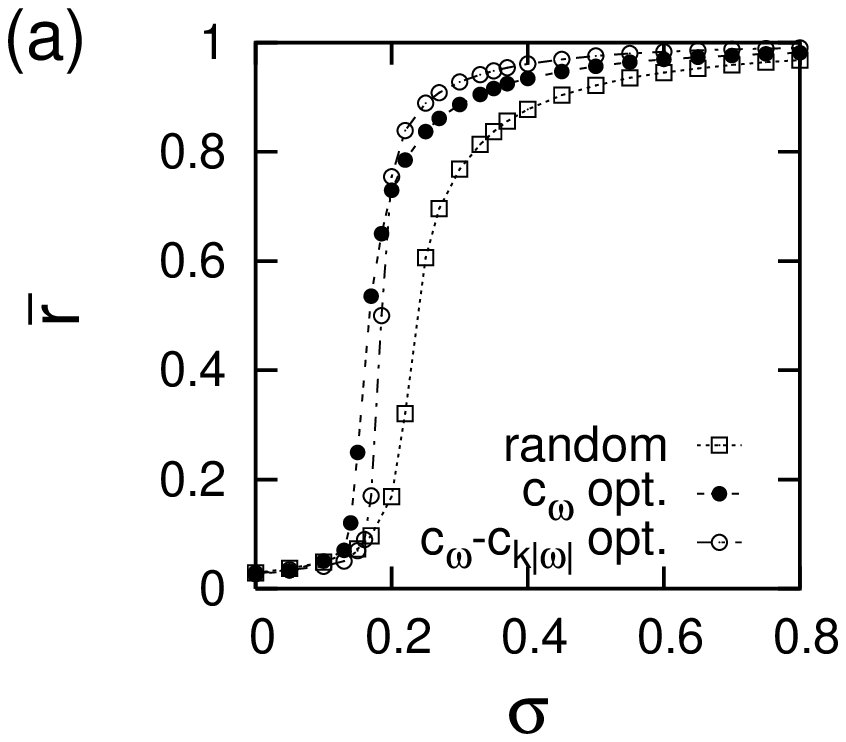}
\includegraphics [width=4.2cm]{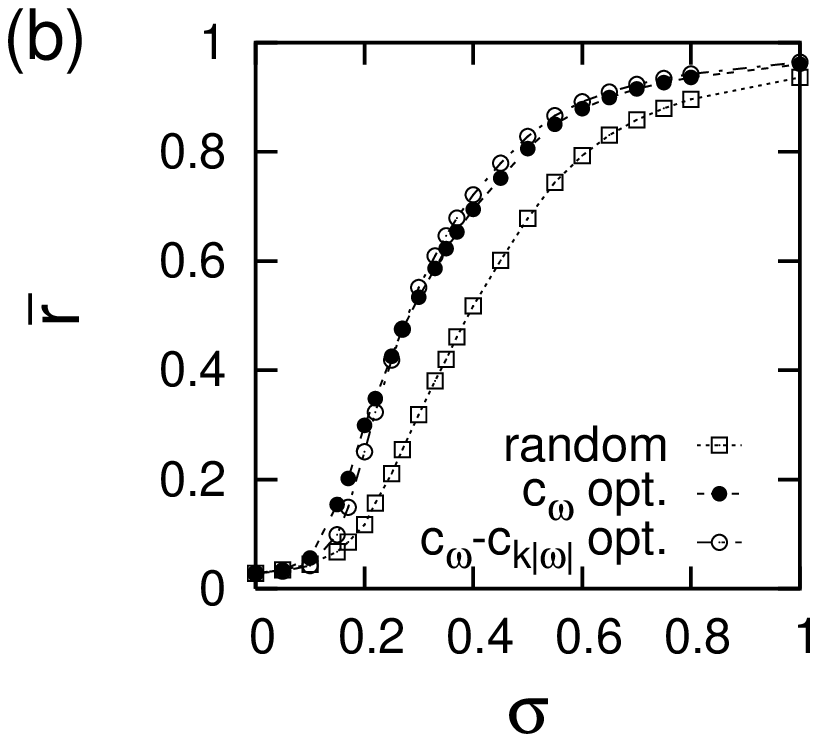}
\caption{Comparison of the synchronizability of (a) ER and (b) scale-free networks with a random distribution of native frequencies (open squares) and distributions with strong anti-correlations between adjacent frequencies (filled circles) and frequency arrangements with an additional correlation between degree and frequency magnitude (open circles). The data are for networks of size $N=1000$ with $\langle k\rangle=6$.}
\label{F3}
\end{center}
\end{figure}

To obtain another independent confirmation for the main contention of the paper, that anti-correlated arrangements of native frequencies on nodes strongly enhance the synchronization behaviour, we proceed by comparing the $r(\sigma)$-dependence of networks with random and strongly anti-correlated $\omega$-assignments. The latter can easily be generated by proposing frequency swaps between nodes on a fixed network and accepting them if they lead to a decrease of $c_\omega$. Figure \ref{F3} shows the dependence of $r$ on $\sigma$ for scale-free and ER networks.
The data corroborate our assertion. It is also quite interesting to generate frequency distributions that additionally positively correlate frequency magnitude and node degree. In the same way as above such an architecture can be created by conditionally swapping frequencies on nodes and minimizing for $c_\omega-c_{|\omega| k}$, where $c_{|\omega| k}$ is a standard correlation coefficient between the $|\omega|$ values and the node degrees $k$. Figure \ref{F3} shows that this frequency arrangement leads to a lower order parameter $r$ for low coupling, but increased $r$ for high $\sigma$'s. This behaviour is straightforward to understand from our previous arguments. The additional correlation of $|\omega|$ and the node degree entails larger differences between native frequencies belonging to a hub and to its average neighbour. As hubs typically serve as germs for clusters of synchrony \cite{Gardenes,Gardenes1}, the latter are harder to form in this scenario.

In summary, we have discussed a method for generating networks of non-identical Kuramoto oscillators with enhanced synchronizability. Using this method, some correlations between properties of the coupling architecture and synchronizability which were previously only derived for the identical oscillator case could be demonstrated for an example system of non-identical oscillators. As another important point, evidence has been given, that the synchronizability depends not only on the topology of the coupling architecture, but is also strongly influenced by the assignement of frequencies to nodes. We have demonstrated that the synchronization behaviour can be strongly enhanced by a frequency arrangement that is marked by strong anti-correlations between frequencies of adjacent nodes. A trade-off between this effect and the requirements for good communication leads to some degree of heterogeneity in the networks with optimal synchronization properties. Further, we have shown that the optimal assignment is such that nodes with more than an average number of connections are assigned native frequencies of large magnitude, while frequencies of small magnitude typically belong to small degree nodes. Our findings give a set of rules for constructing more synchronizable networks of heterogeneous oscillators. These might be of practical importance for engineering problems, but also give predictions that should be tested for a range of biological networks where synchronization plays a role.


\begin{thebibliography}{99}
\bibitem {remark0} {We also experimented with other distributions of frequencies, e.g., Normal and bimodal $\rho(\omega)=1/2(\delta(\omega+1)+\delta(\omega-1))$, but found the main results to be independent of this choice.}
\end{thebibliography}
\end{document}